\documentclass{fuin}
\usepackage{amsmath, amssymb}
\usepackage{graphics}
\usepackage{epsfig}

\begin{document}

\newcommand{\cell}[1]{\makebox[0.25cm]{#1}}

\title{Number-conserving cellular automaton rules}

\author{Nino Boccara\dag\ and Henryk Fuk\'s\ddag\\
\dag\ Department of Physics, University of Illinois, Chicago, USA\\
\texttt{boccara@uic.edu}\\
and\\
DRECAM/SPEC, CE Saclay, 91191 Gif-sur-Yvette Cedex, France\\
\ddag\ The Fields Institute for Research in Mathematical
Sciences,\\
Toronto ON M5T 2W1, Canada\\
\texttt{hfuks@fields.utoronto.ca}
and\\
Department of Mathematics and Statistics, University of Guelph, \\
Guelph, ON N1G 2W1, Canada}

\maketitle


\runninghead{N. Boccara and H. Fuk\'s}{Number-conserving CA rules}

\abstract{A necessary and sufficient condition for a one-dimen\-sional
$q$-state $n$-input cellular automaton rule to be number-conserving
is established. Two different forms of simpler and more visual
representations of these rules are given, and their flow diagrams are
determined. Various examples are presented and applications to car
traffic are indicated. Two nontrivial three-state three-input
self-conjugate rules have been found. They can be used to model the
dynamics of random walkers.}

\section{Introduction}

A one-dimensional cellular automaton (CA) is a discrete
dynamical system, which may be defined as follows. Let
$s:\mathbb{Z}\times\mathbb{N}\mapsto\mathcal{Q}$ be a function that
satisfies the equation
\begin{equation}
s(i,t+1)=f\big(s(i-r_\ell,t),s(i-r_\ell+1,t),\ldots,s(i+r_r,t)\big),
\label{evolution}
\end{equation}
for all $i\in\mathbb{Z}$, and all $t\in\mathbb{N}$. $\mathbb{Z}$
is the set of all integers, $\mathbb{N}$ the set of nonnegative
integers, and $\mathcal{Q}$ a finite set of states, usually
equal to $\{0,1,2,\ldots,q-1\}$. $s(i,t)$ represents the
\textit{state of site $i$ at time $t$}, and the mapping
$f : {\mathcal{Q}}^{r_\ell+r_r+1}\to\mathcal{Q}$ is the \textit{CA evolution
rule}. The positive integers $r_\ell$ and $r_r$ are, respectively,
the \textit{left} and \textit{right radii} of the
rule. In what follows, $f$ will be referred to as an \textit{$n$-input rule},
where $n$ is the number $r_\ell+r_r+1$ of arguments of $f$. Following
Wolfram \cite{wolf}, to each rule $f$ we assign a \textit{rule number}
$N(f)$ such that
$$
N(f) = \sum_{(x_1,x_2,\ldots,x_n)\in{\mathcal{Q}}^n}
f(x_1,x_2,\ldots,x_n)q^{q^{n-1}x_1+q^{n-2}x_2+\cdots+q^0x_n}.
$$
Cellular automata (CAs) have been widely used to model complex systems in
which the local character of the evolution rule plays an essential
role \cite{PBVB,gut,BGMP,BSS}. In this paper, we will only consider
finite CAs, and replace
the set $\mathbb{Z}$ by the set ${\mathbb{Z}}_L$ of integers modulo $L$. Any
element of the set ${\mathcal{Q}}^L$ will be called an
\textit{$L$-cycle} or a \textit{cyclic configuration of length $L$}.

Recently, following Nagel and Schreckenberg \cite{NS}, many authors proposed
various CA models of highway traffic flow. In the simplest
models of this type, ${\mathbb{Z}}_L$ represents a one-lane circular
highway, and ${\mathcal{Q}}$ is equal to $\{0,1\}$.
According to the value of $s(i,t)$, it is said that, at time $t$, site $i$
is either empty, if $s(i,t)=0$, or occupied by a car, if $s(i,t)=1$.
In order to complete the description of the model, an evolution
rule has to be defined. For the sake of simplicity, we might assume that,
at each time step, all cars move to the right neighboring site if, and only if,
this site is empty. It is not difficult to verify that this rule coincides
with Rule 184 defined by
\begin{alignat*}{4}
f_{184}(0,0,0) & = 0, &\quad f_{184}(0,0,1) & = 0, &\quad f_{184}(0,1,0) & = 0,
&\quad f_{184}(0,1,1) & = 1,\\
f_{184}(1,0,0) & = 1, &\quad f_{184}(1,0,1) & = 1, &\quad f_{184}(1,1,0) & = 0,
&\quad f_{184}(1,1,1) & = 1.
\end{alignat*}
This traffic rule does not allow cars to enter or exit the highway. Therefore,
during the evolution, the number of cars should remain constant, that is,
starting from any initial cyclic configuration of length $L$, and for all
$t\in\mathbb{N}$, Rule 184 should satisfy the condition
\begin{multline}
f_{184}\big(s(1,t),s(2,t),s(3,t)\big) + f_{184}\big(s(2,t),s(3,t),s(4,t)\big)
+ \cdots \\
+ f_{184}\big(s(L,t),s(1,t),s(2,t)\big) = s(1,t) + s(2,t) + \cdots + s(L,t).
\label{184-conserve}
\end{multline}
Since Rule 184 may also be written
\begin{equation}
f_{184}(x_1,x_2,x_3) = x_2 + \min\{x_1, 1-x_2\} - \min\{x_2,1-x_3\},
\label{CR184}
\end{equation}
condition~(\ref{184-conserve}) is clearly verified.
If we had assumed that cars, instead of moving to the right, had to move to
the left, we would have found that the evolution rule would have
been Rule 226. The relations
\begin{alignat*}{2}
f_{226}(x_1,x_2,x_3) & = f_{184}(x_3,x_2,x_1)\\
f_{226}(x_1,x_2,x_3) & = 1- f_{184}(1-x_1,1-x_2,1-x_3),
\end{alignat*}
show that these two rules are not fundamentally distinct.

Rules 184 and 226 are the simplest nontrivial examples of number-conserving
CAs. In a previous paper \cite{BF}, in order to find all the deterministic car
traffic rules, we determined, up to $n=5$, all the 2-state number-conserving
CA rules. This was done using a different technique than the one presented in this
paper. For $n>3$, not all these rules did correspond to realistic car traffic
rules since they allowed vehicles to move in both directions. We found then
that it was more appropriate to interpret all these 2-state number-conserving
CA rules as evolution operators of one-dimensional systems of distinguishable
particles.

The purpose of this paper is to derive a necessary and sufficient condition
for a one-dimensional $q$-state $n$-input CA rule to be number-conserving.
We will then give examples of such rules and indicate some of their applications.
Our result is an illustration of a general theorem on additive conserved
quantities established by Hattori and Takesue \cite{HT}.

\section{Number-conserving rules}

\definition{A one-dimensional $q$-state $n$-input CA rule $f$
is \textit{number-con\-serving} if, for all cyclic configurations of length
$L\ge n$, it satisfies
\begin{multline}
f(x_1,x_2,\ldots,x_{n-1},x_n)+f(x_2,x_3,\ldots,x_n,x_{n+1})+\cdots\\
+f(x_L,x_1\ldots,x_{n-2},x_{n-1})=x_1+x_2+\cdots+x_L.
\label{CRf}
\end{multline}}

\begin{theorem}A one-dimensional $q$-state $n$-input CA rule $f$ is
number-con\-serving if, and only if, for all
$(x_1,x_2,\ldots,x_n)\in{\mathcal{Q}}^n$, it satisfies
\begin{align}
f(x_1,x_2,\ldots,x_n) = x_1 + \sum_{k=1}^{n-1}\big(
&f(\underbrace{0,0,\ldots,0}_k,x_2,x_3,\ldots,x_{n-k+1})\notag\\
-&f(\underbrace{0,0,\ldots,0}_k,x_1,x_2,\ldots,x_{n-k})\big).
\label{NScond}
\end{align}
\end{theorem}

To simplify the proof we will need the following lemma.

\begin{lemma}If $f$ is a number-conserving rule, then
\begin{equation}
f(0,0,\ldots,0)=0.
\label{f0}
\end{equation}
\end{lemma}

Write Condition~(\ref{CRf}) for a cyclic configuration of length $L\ge n$
whose all elements are equal to zero.\hfill$\square$

\vspace{0.3cm}

To prove that Condition (\ref{NScond}) is necessary, consider a cyclic
configuration of length $L\ge 2n-1$ which is the
concatenation of a sequence $(x_1,x_2,\ldots,x_n)$ and a
sequence of $L-n$ zeros, and express that the $n$-input rule $f$ is
number-conserving. We obtain
\begin{align}
f(0,0,\ldots,0,x_1)& + f(0,0,\ldots,0,x_1,x_2) + \cdots\notag\\
&\qquad + f(x_1,x_2,\ldots,x_n) + f(x_2,x_3,\ldots,x_n,0)+\cdots\notag\\
&\qquad + f(x_n,0,\ldots,0) = x_1+x_2+\cdots+x_n,
\label{x1neq0}
\end{align}
where all the terms of the form $f(0,0,\ldots,0)$, which are equal
to zero according to (\ref{f0}), have not been written.
Replacing $x_1$ by 0 in (\ref{x1neq0}) gives
\begin{align}
f(0,0,\ldots,0,x_2)& + \cdots + f(0,x_2,\ldots,x_n)\notag\\
& + f(x_2,x_3,\ldots,x_n,0) + \cdots + f(x_n,0,\ldots,0)\notag\\
& = x_2+\cdots+x_n.
\label{x1eq0}
\end{align}
Subtracting (\ref{x1eq0}) from (\ref{x1neq0}) yields
(\ref{NScond}).

Condition (\ref{NScond}) is obviously sufficient since, when summed
on a cyclic configuration, all the left-hand side terms except the
first cancel.\hfill$\square$

\begin{remark}The above proof shows that if we can verify that a CA rule $f$ is
number-conserving for all cyclic configurations of length $2n-1$, then it is
number-conserving for all cyclic configurations of length $L>2n-1$.
\end{remark}

The following corollaries are simple necessary conditions for a CA
rule to be number-conserving.

\begin{corollary}If $f$ is a one-dimensional $q$-state $n$-input
number-conserving CA rule, then, for all $x\in\mathcal{Q}$,
\begin{equation}
f(x,x,\ldots,x)= x.
\label{fx}
\end{equation}
\end{corollary}

To prove~(\ref{fx}), which is a generalization of (\ref{f0}), write
Condition (\ref{NScond}) for $x_1=x_2=\cdots=x_n=x$.\hfill$\square$

\begin{corollary}If $f$ is a one-dimensional $q$-state $n$-input
number-conserving CA rule, then,
\begin{equation}
\sum_{(x_1,x_2,\ldots,x_n)\in{\mathcal{Q}}^n}
f(x_1,x_2,\ldots,x_n) = \frac{1}{2}(q-1)\,q^n.
\label{sumf}
\end{equation}
\end{corollary}

When we the sum (\ref{NScond}) over $(x_1,x_2,\ldots,x_n)\in{\mathcal{Q}}^n$,
all the left-hand side terms except the first cancel, and the sum over the
remaining term is equal to $(0+1+2+\cdots+(q-1))q^{n-1}=\frac{1}{2}(q-1)\,q^n$.
\hfill$\square$

It is possible to give an interesting alternative proof of
Relation~(\ref{sumf}). Consider a De Bruijn cycle of length $q^n$.
Such a cycle contains all the different $n$-tuples
$(x_1,x_2,\ldots,x_n)\in{\mathcal{Q}}^n$. The existence of De Bruijn
cycles is related to the existence of Eulerian circuits on a De Bruijn
graph $G_{n-1}$. Such a graph has $q^{n-1}$ vertices and $q^n$ arcs.
The number of arcs leaving a vertex is the same as the number which arrive.
Each vertex is labeled by an $(n-1)$-tuple over $\mathcal{Q}$, and the
arc joining the vertex $(x_1,x_2,\ldots,x_{n-1})$ to the vertex
$(x_2,\ldots,x_{n-1},x_n)$ is labeled $(x_1,x_2,\ldots,x_{n-1},x_n)$.
For example,
$$
(0,0,0,0,1,0,1,0,0,1,1,0,1,1,1,1)
$$
is a De Bruijn cycle for $q=2$ and $n=4$. It can be shown that there exist
$q^{-n}(q\,!)^{q^{n-1}}$ distinct De Bruijn cycles \cite{lint}.
Since all the elements of $\mathcal{Q}$ appear an equal number of
times in a De Bruijn cycle, Condition~(\ref{CRf}) implies
$$
\sum_{(x_1,x_2,\ldots,x_n)\in{\mathcal{Q}}^n}
f(x_1,x_2,\ldots,x_n) = \big(0+1+2+\cdots+(q-1)\big)\,q^{n-1},
$$
and (\ref{sumf}) follows.\hfill$\square$

\section{Examples of number-conserving rules}

Using Condition~(\ref{NScond}) we can determine all the
number-conserving CA rules for fixed values of $q$ and $n$. These rules will be
considered as operators governing the discrete dynamics of systems of
particles whose total number is conserved. The particles occupy the cells
of a one-dimensional periodic lattice subject to the condition that, at a given
time, no more than $q-1$ particles may occupy the same cell. Among these
rules some have similar properties, which can be exhibited using
the operators of reflection and conjugation. These two operators denoted,
respectively by $R$ and $C$, are defined on the set of all
one-dimensional $q$-state $n$-input cellular automaton rules by
\begin{align*}
R\,f(x_1,x_2,\ldots,x_n) & = f(x_n,x_{n-1},\ldots,x_1)\\
C\,f(x_1,x_2,\ldots,x_n) & = q-1 - f(q-1-x_1, q-1-x_2,\ldots,q-1-x_n).
\end{align*}
It is clear that, if $f$ is number-conserving, then $R\,f$,
$C\,f$, and $RC\,f=CR\,f$ have the same property. Rules $f$ and
$R\,f$ govern identical dynamics, the only difference is that, if
for one rule particles flow to the right, for the other rule, they will
flow in the opposite direction. To understand the difference
between rules $f$ and $C\,f$, let us assume that a cell, which contains
$k$ particles ($1\le k\le q-1$) contains $q-1-k$ holes. Then,
conjugation may be viewed as exchanging the roles of particles and
holes. That is, if $f$ describes a specific motion of particles then
$C\,f$ describes the same rule, but for the motion of holes.

When the number of states and number of inputs are not very small, the
dynamics of the particles is not clearly exhibited by the rule table of
a number-conserving CA rule. A simpler and more visual picture of the
rule is given by its \textit{motion representation}. Such a motion
representation may be defined as follows. List all the neighborhoods of
a given occupied site represented by its site value $\mathbf{s}$. Then,
for each neighborhood, indicate the displacements of the $s$
particles by a nondecreasing sequence of $s$ integers $(v_1,v_2,\ldots,v_s)$
representing the different velocities of the $s$ particles. Velocities are
positive if particles move to the right and negative if they move to the left.
Alternatively, to the sequence $(v_1,v_2,\ldots,v_s)$, we can
substitute arrow(s) joining the site where are initially located the s
particles to the final positions of the particle(s). A number above the
arrow indicates how many particles are moving to this final position.
To simplify a bit more these representations, we only list
neighborhoods for which, at least one velocity $v_j$ ($j=1,2,\ldots,s$)
is different from zero. For example, these two forms of the motion
representation of 2-state 3-input Rule 184 are
$$
{\mathbf{1}0}\ \ (1)\qquad\text{and}\qquad
\overset{\overset{1}{\curvearrowright}}{10}.
$$
Since, for Rule 184, particles move only to the right, there is no need
to indicate the state of the left neighboring site of the particle.
Many other examples of motion representations for 2-state 4- and
5-input rules are given in~\cite{BF}.

\begin{remark}In many applications, as one-way road car traffic,
which prohibits passing, we have to assume that the particles are
distinguishable. We have, therefore, to label them using an
increasing sequence of integers, in order to be able to follow the
motion of each individual particles. Instead of describing
formally the labeling process, we will give a simple example.
Suppose we have the configuration:
\begin{center}
\begin{tabular}{|c|c|c|c|c|c|c|c|c|}
\hline
\cell{$\cdots$}&\cell{0}&\cell{1}&\cell{0}&\cell{2}&\cell{0}&\cell{2}
&\cell{0}&\cell{$\cdots$}\\
\hline
\end{tabular}
\end{center}
We first replace each particle in each occupied cell by the symbol
$\bullet$
\begin{center}
\begin{tabular}{|c|c|c|c|c|c|c|c|c|}
\hline
\cell{$\cdots$}&\cell{}&\cell{$\bullet$}&\cell{}&\cell{$\bullet\ \bullet$}
&\cell{}&\cell{$\bullet\ \bullet$}&\cell{}&\cell{$\cdots$}\\
\hline
\end{tabular}
\end{center}
and then label consecutively all the $\bullet$ to obtain:
\begin{center}
\begin{tabular}{|c|c|c|c|c|c|c|c|c|}
\hline
\cell{$\cdots$}&\cell{}&\cell{1}&\cell{}&\cell{2\ 3}&\cell{}&\cell{4\ 5}
&\cell{}&\cell{$\cdots$}\\
\hline
\end{tabular}
\end{center}
If now, we assume that the motion representation is
$$
0\mathbf{1}\ (-1)\quad
0\mathbf{2}0\ (-1,1)\quad
0\mathbf{2}1\ (-1,1)\quad
1\mathbf{2}0\ (0,1)\quad
1\mathbf{2}1\ (0,1)
$$
$$
2\mathbf{2}0\ (0,1)\quad
2\mathbf{2}1\ (0,1)\quad
0\mathbf{2}2\ (-1,0)\quad
0\mathbf{2}2\ (-1,0),
$$
then, applying this rule, we obtain the new labeled configuration:
\begin{center}
\begin{tabular}{|c|c|c|c|c|c|c|c|c|}
\hline
\cell{$\cdots$}&\cell{1}&\cell{}&\cell{2}&\cell{}&\cell{3\ 4}&\cell{}
&\cell{5}&\cell{$\cdots$}\\
\hline
\end{tabular}
\end{center}
which corresponds to the new configuration:
\begin{center}
\begin{tabular}{|c|c|c|c|c|c|c|c|c|}
\hline
\cell{$\cdots$}&\cell{1}&\cell{0}&\cell{1}&\cell{0}&\cell{2}&\cell{0}
&\cell{1}&\cell{$\cdots$}\\
\hline
\end{tabular}
\end{center}
Note that the above labeling convention assumes that particles
cannot jump above each other, and, therefore, the ordering of labels remains
unchanged after each iteration. We use this convention to ensure
uniqueness of the motion representation for number-conserving CA rules.
\end{remark}

\subsection{Three-state two-input number-conserving rules}

There are only 4 three-state two-input number-conserving rules,
which are listed in Table~1.

For radii pair $(r_\ell, r_r)$, we have two possible choices,
either $(0,1)$ or $(1,0)$. In each case, we have chosen
the pair simplifying most the motion representation.
Rules 19305 and 15897 coincide with the identity.
The motion representations of Rules 18561 and 16641, which are obtained
from one another by reflection, are, respectively,
\begin{equation*}
{\mathbf{2}0}\ \ (0,1),\quad {\mathbf{2}1}\ \ (0,1)\ \
\text{and}\ \
{0\mathbf{2}}\ \ (-1,0),\quad {1\mathbf{2}}\ \ (-1,0),
\end{equation*}
or
\begin{equation*}
\overset{\overset{1}{\curvearrowright}}{20},\quad
\overset{\overset{1}{\curvearrowright}}{21},\quad
\text{and}\quad
\overset{\overset{1}{\curvearrowleft}}{02},\quad
\overset{\overset{1}{\curvearrowleft}}{12}.
\end{equation*}
For both rules, if a site is occupied by 2 particles, one of them move to
a neighboring site, except if this site is already occupied by 2 particles.

\begin{table}
\begin{center}
\begin{tabular}[htb]{|c|c|c|}
\hline
rule number & base 3 representation & $(r_\ell,r_r)$ \\
\hline\hline
19305 & 222111000 & (0,1)\\
15897 & 210210210 & (1,0)\\
18561 & 221110110 & (1,0)\\
16641 & 211211100 & (0,1)\\
\hline
\end{tabular}
\end{center}
\caption{Three-state two-input minimal number-conserving rules.}
\end{table}

\subsection{Three-state three-input number-conserving rules}

There are 144 three-state three-input number-conserving rules.
They can be divided into 48 equivalence classes under the dihedral group
generated by the operators $R$ and $C$. Each class will be represented by
the rule having the smallest rule number, and which will be called the minimal
rule of the class. Tables~2 and 3 list
rule numbers and corresponding base-3 representations of all 48 minimal rules.

\begin{table}
\begin{center}
\begin{tabular}[hbt]{|c|c|c|}
\hline
reference number & rule number & base-3 representation\\
\hline\hline
1  & 6159136430181 &  210210210210210210210210210\\
2  & 6159523870341 &  210210211210210211210210100\\
3  & 6169984499133 &  210211211210211100210211100\\
4  & 6171146819301 &  210211221210211110210100110\\
5  & 6171534259461 &  210211222210211111210100000\\
6  & 6181994888253 &  210212222210212000210101000\\
7  & 6201360244413 &  210221211210110100210221100\\
8  & 6202522564581 &  210221221210110110210110110\\
9  & 6202910004741 &  210221222210110111210110000\\
10 & 6213370633533 &  210222222210111000210111000\\
11 & 6436931290101 &  211210100211210211211210100\\
12 & 6447391918893 &  211211100211211100211211100\\
13 & 6448554239061 &  211211110211211110211100110\\
14 & 6448941679221 &  211211111211211111211100000\\
15 & 6450878870973 &  211211200211211200100211200\\
16 & 6452428631301 &  211211211211211211100100100\\
17 & 6459402308013 &  211212111211212000211101000\\
18 & 6462889260093 &  211212211211212100100101100\\
19 & 6478767664173 &  211221100211110100211221100\\
20 & 6480317424501 &  211221111211110111211110000\\
21 & 6482254616253 &  211221200211110200100221200\\
22 & 6483804376581 &  211221211211110211100110100\\
23 & 6490778053293 &  211222111211111000211111000\\
24 & 6491940373461 &  211222121211111010211000010\\
\hline
\end{tabular}
\end{center}
\caption{Three-state three-input minimal number-conserving rules, Part 1.}
\end{table}

\begin{table}
\begin{center}
\begin{tabular}[hbt]{|c|c|c|}
\hline
reference number & rule number & base-3 representation\\
\hline\hline
25 & 6494265005373 &  211222211211111100100111100\\
26 & 6495427325541 &  211222221211111110100000110\\
27 & 6726349098981 &  212211000212211111212100000\\
28 & 6729836051061 &  212211100212211211101100100\\
29 & 6736809727773 &  212212000212212000212101000\\
30 & 6740296679853 &  212212100212212100101101100\\
31 & 6757724844261 &  212221000212110111212110000\\
32 & 6761211796341 &  212221100212110211101110100\\
33 & 6768185473053 &  212222000212111000212111000\\
34 & 6769347793221 &  212222010212111010212000010\\
35 & 6771672425133 &  212222100212111100101111100\\
36 & 6881331565845 &  220100211220211100220211100\\
37 & 6893341954965 &  220101222220212000220101000\\
38 & 6912707311125 &  220110211220110100220221100\\
39 & 6924717700245 &  220111222220111000220111000\\
40 & 6956093445525 &  220121222220010000220121000\\
41 & 7158738985605 &  221100100221211100221211100\\
42 & 7170749374725 &  221101111221212000221101000\\
43 & 7174236326805 &  221101211221212100110101100\\
44 & 7202125120005 &  221111111221111000221111000\\
45 & 7205612072085 &  221111211221111100110111100\\
46 & 7233500865285 &  221121111221010000221121000\\
47 & 7448156794485 &  222101000222212000222101000\\
48 & 7479532539765 &  222111000222111000222111000\\
\hline
\end{tabular}
\end{center}
\caption{Three-state three-input minimal number-conserving rules, Part 2.}
\end{table}

Among these minimal rules, 18 eventually emulate the identity. Their
reference numbers are: 27, 28, 29, 30, 31, 32, 33, 35, 37, 39, 40, 42,
43, 44, 45, 46, 47, 48. The motion representations of the remaining 30
rules are given in Table~4.

\begin{table}
\begin{center}
\begin{tabular}[hbt]{|c|c|}
\hline
reference number & motion representation\\
\hline\hline
1  & \textbf{1}\ (-1)\quad
\textbf{2}\ (-1,-1)\\
2  & \textbf{1}0\ (1)\quad
\textbf{2}0\ (-1,1)\quad
\textbf{2}1\ (-1,0)
\textbf{2}2\ (-1,0)\\
3  & \textbf{2}0\ (-1,1)\quad
\textbf{2}1\ (-1,0)\quad
\textbf{2}2\ (-1,0)\\
4  & 0\textbf{1}\ (-1)\quad
\textbf{2}0\ (-1,1)\quad
\textbf{2}1\ (-1,0)\quad
\textbf{2}2\ (-1,0)\\
5  & \textbf{1}0\ (1)\quad
\textbf{1}1\ (1)\quad
\textbf{2}0\ (1,1)\quad
\textbf{2}1\ (0,1)\\
6  & \textbf{2}0\ (1,1)\quad
\textbf{1}1\ (1)\quad
\textbf{2}1\ (0,1)\\
7  & \textbf{1}1\ (-1)\quad
\textbf{2}0\ (-1,1)
\textbf{2}1\ (-1,0)\quad
\textbf{2}2\ (-1,0)\\
8  & 0\textbf{1}\ (-1)\quad
1\textbf{1}\ (-1)\quad
\textbf{2}0\ (-1,1)\quad
\textbf{2}1\ (-1,0)\quad
\textbf{2}2\ (-1,0)\\
9  & \textbf{1}0\ (1)\quad
\textbf{2}0\ (1,1)\quad
\textbf{2}1\ (0,1)\\
10  & \textbf{2}0\ (1,1)\quad
\textbf{2}1\ (0,1)\\
11  & \textbf{1}0\ (1)\quad
\textbf{2}\ (-1,0)\\
12  & \textbf{2}\ (-1,0)\\
13  & 0\textbf{1}\ (-1)\quad
\textbf{2}\ (-1,0)\\
14  & \textbf{1}0\ (1)\quad
\textbf{1}1\ (1)\quad
\textbf{2}0\ (0,1)\quad
\textbf{2}1\ (0,1)\\
15  & 0\textbf{2}\ (-1,-1)\quad
1\textbf{2}\ (-1,0)\quad
2\textbf{2}\ (-1,0)\\
16  & \textbf{1}0\ (1)\quad
\textbf{1}1\ (1)\quad
0\textbf{2}0\ (-1,1)\quad
0\textbf{2}1\ (-1,1)\quad
1\textbf{2}0\ (0,1)\\
    & 1\textbf{2}1\ (0,1)\quad
2\textbf{2}0\ (0,1)\quad
2\textbf{2}1\ (0,1)\quad
0\textbf{2}2\ (-1,0)\\
17  & \textbf{1}1\ (1)\quad
\textbf{2}0\ (0,1)\quad
\textbf{2}1\ (0,1)\\
18  & 0\textbf{2}0\ (-1,1)\quad
0\textbf{2}1\ (-1,1)\quad
1\textbf{2}0\ (0,1)\quad
1\textbf{2}1\ (0,1)\\
    & 2\textbf{2}0\ (0,1)\quad
2\textbf{2}1\ (0,1)\quad
\textbf{1}1\ (1)\quad
0\textbf{2}2\ (-1,0)\\
19  & 1\textbf{1}\ (-1)\quad
\textbf{2}\ (-1,0)\\
20  & \textbf{1}0\ (1)\quad
\textbf{2}0\ (0,1)\quad
\textbf{2}1\ (0,1)\\
21  & 1\textbf{1}\ (-1)\quad
0\textbf{2}\ (-1,-1)\quad
1\textbf{2}\ (-1,0)\quad
2\textbf{2}\ (-1,0)\\
22   & \textbf{1}0\ (1)\quad
0\textbf{2}0\ (-1,1)\quad
0\textbf{2}1\ (-1,1)\quad
1\textbf{2}0\ (0,1)\quad
1\textbf{2}1\ (0,1)\\
     & 2\textbf{2}0\ (0,1)\quad
2\textbf{2}1\ (0,1)\quad
0\textbf{2}2\ (-1,0)\\
23  & \textbf{2}0\ (0,1)\quad
\textbf{2}1\ (0,1)\\
24  & 0\textbf{1}\ (-1)\quad
\textbf{2}0\ (0,1)\quad
\textbf{2}1\ (0,1)\\
25   & 0\textbf{2}0\ (-1,1)\quad
0\textbf{2}1\ (-1,1)\quad
1\textbf{2}0\ (0,1)\quad
1\textbf{2}1\ (0,1)\\
     & 2\textbf{2}0\ (0,1)\quad
2\textbf{2}1\ (0,1)\quad
0\textbf{2}2\ (-1,0)\\
26   & 0\textbf{1}\ (-1)\quad
0\textbf{2}0\ (-1,1)\quad
0\textbf{2}1\ (-1,1)\quad
1\textbf{2}0\ (0,1)\quad
1\textbf{2}1\ (0,1)\\
     & 2\textbf{2}0\ (0,1)\quad
2\textbf{2}1\ (0,1)\quad
0\textbf{2}2\ (-1,0)\quad
1\textbf{2}2\ (-1,0)\\
34  & 0\textbf{1}\ (-1)\quad
\textbf{2}1\ (0,1)\\
36  & 2\textbf{1}\ (-1)\quad
\textbf{2}0\ (-1,1)\quad
\textbf{2}1\ (-1,0)\quad
\textbf{2}2\ (-1,0)\\
38  & 1\textbf{1}\ (-1)\quad
2\textbf{1}\ (-1)\quad
\textbf{2}0\ (-1,1)\quad
\textbf{2}1\ (-1,0)\quad
\textbf{2}2\ (-1,0)\\
41  & 2\textbf{1}\ (-1)\quad
\textbf{2}\ (-1,0)\\
\hline
\end{tabular}
\end{center}
\caption{Motion representations of the 30 minimal which do not emulate
the identity.}
\end{table}

\begin{figure}[hbt]
\centerline{\includegraphics*[scale=0.7]{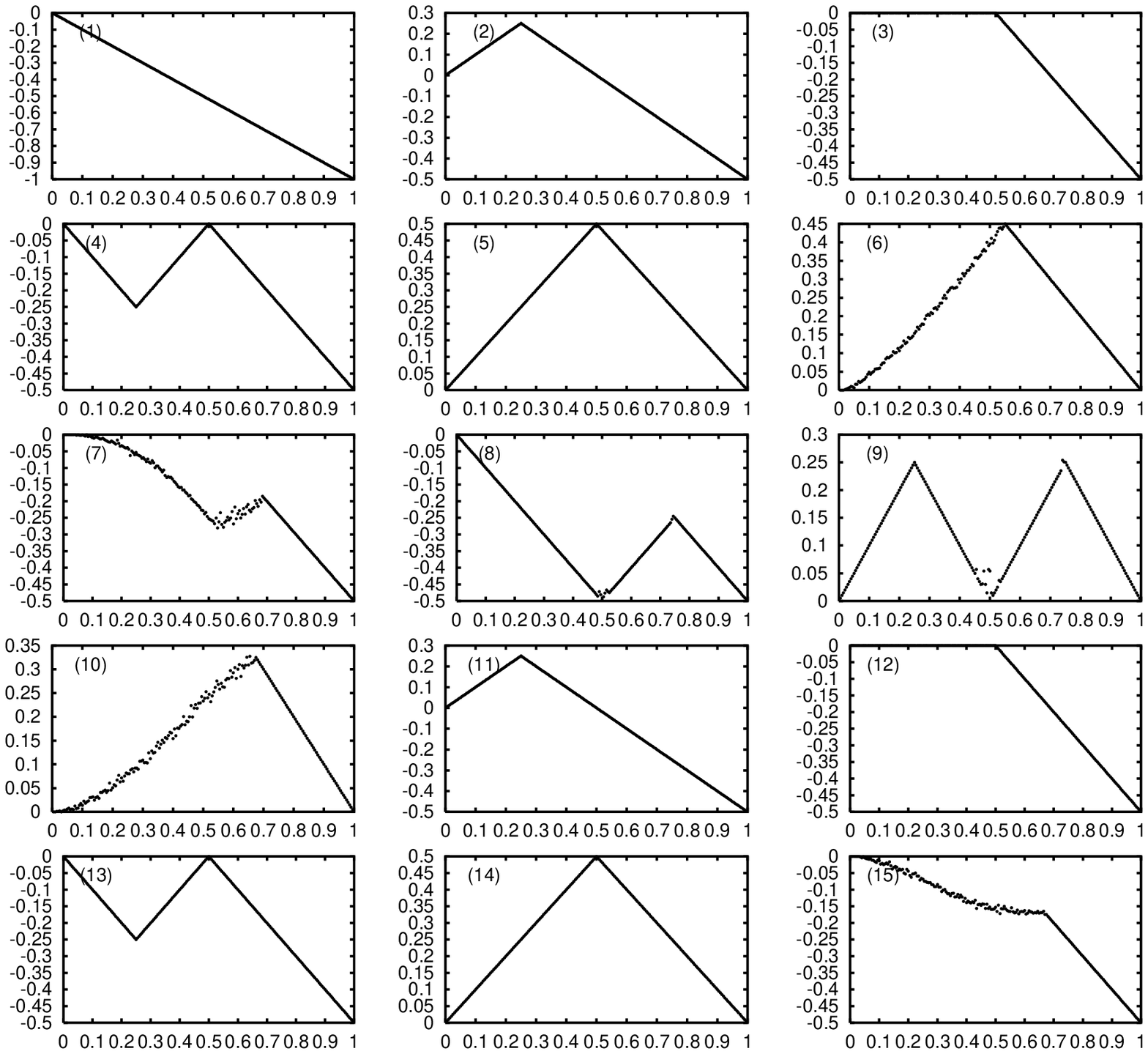}}
\caption{\label{fig:flow1}\textit{Flow diagrams, Part 1.}}
\end{figure}

\begin{figure}[hbt]
\centerline{\includegraphics*[scale=0.7]{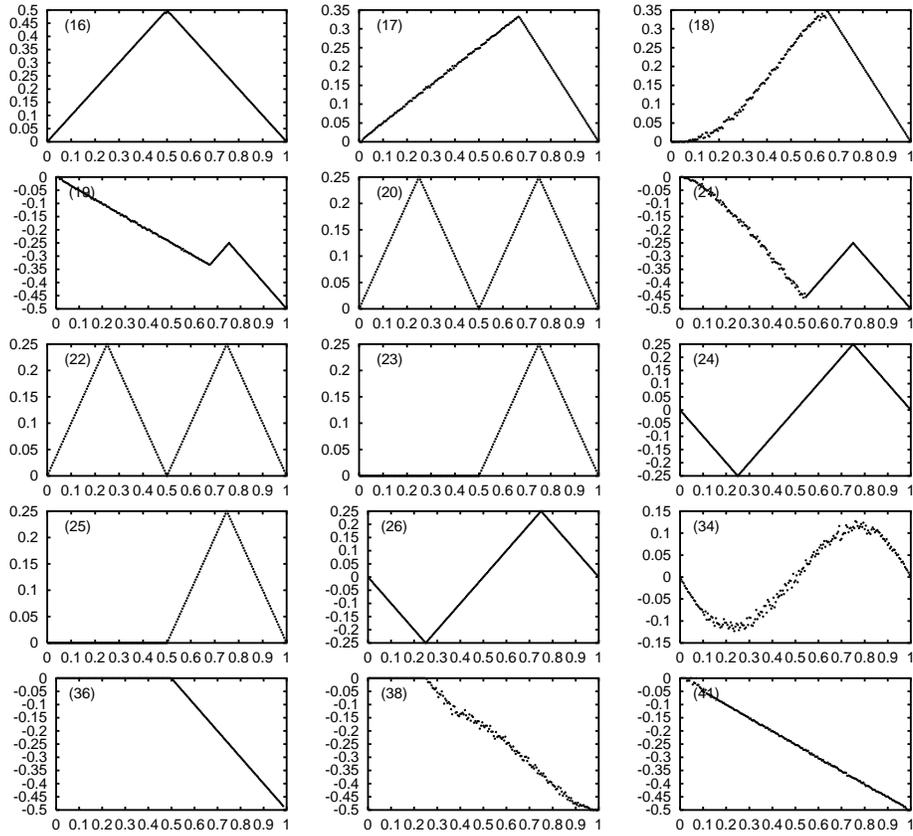}}
\caption{\label{fig:flow2}\textit{Flow diagrams, Part 2.}}
\end{figure}

In order to have a global view of the properties of the various
minimal rules, Figures~\ref{fig:flow1} and~\ref{fig:flow2} represent
the flow diagrams of the 30 minimal rules whose motion representations are
listed in Table~4. If $\rho$ is the average particles
density and $v_{\rm av}$ the average particles velocity, a flow diagram shows
how the flow $\rho v_{\rm av}$ varies as a function of $\rho$. They are
common practice in car traffic theory. Concerning these flow diagrams, one point
has to be stressed. For $q>2$, that is, if more than one particle
can occupy a site, it is not always possible to choose the left and right
radii of a rule so as to have $v_{\rm av}=0$ for $\rho=1$. This is due to the
fact that, choosing $(r_\ell,r_r)=(1,1)$, for some rules, $\rho v_{\rm av}=-0.5$
when $\rho=1$. Any other choice will either increase or decrease $v_{\rm av}$
by one unit. The flow of the rules which emulate the identity, is equal to
zero for all values of $\rho$.

\begin{remark}All the rules for which the flow is non-negative for all
values of the average density, could be \textit{a priori} considered as
deterministic one-lane highway car traffic. Consider for instance
Rule 6171534259461 (reference number 5), the second form of its motion
representation is
$$
\overset{\overset{1}{\curvearrowright}}{10}\quad
\overset{\overset{1}{\curvearrowright}}{11}\quad
\overset{\overset{2}{\curvearrowright}}{20}\quad
\overset{\overset{1}{\curvearrowright}}{21}.
$$
That is, as many particles as possible move from one site to the right
neighboring site. This rule, which is the simplest three-state rule
generalizing Rule 184, may be written
$$
f(x_1,x_2,x_3) = x_2 + \min\{x_1,2-x_2\} - \min\{x_2,2-x_3\}.
$$
This expression suggests a further generalization. The $q$-state rule
defined by, for all $(x_1,x_2,x_3)\in\{0,1,\ldots,q-1\}^3$,
$$
f(x_1,x_2,x_3) = x_2 + \min\{x_1,q-1-x_2\} - \min\{x_2,q-1-x_3\},
$$
can be viewed as the following car traffic rule: Each cell
represents a section of a one-way road between two traffic lights. The
number of states $q$ measures the maximum capacity of that
section. The above evolution rule describes the way cars move
from one section to the next. As for elementary CA Rule 184, the
flow is maximal for $\rho=0.5$. Of course, and this goes beyond
our purpose here, we could, as in the  Nagel--Schreckenberg
model \cite{NS}, introduce some noise and say that some cars, which
could move to the next section, do not do so with a probability $p$.
The existence of traffic lights could be used to monitor traffic
in order to increase the flow, and, with this in mind, we could
extend to $q$-state rules classes of models we studied in a recent
paper~\cite{FB}.
\end{remark}

\begin{figure}[hbt]
\centerline{\includegraphics*[scale=0.6]{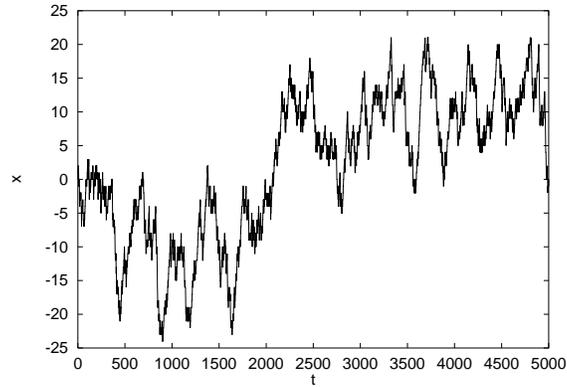}}
\caption{\label{fig:trace26}\textit{Random walker evolving according to Rule
6495427325541 (no 26).}}
\end{figure}

\begin{figure}[hbt]
\centerline{\includegraphics*[scale=0.6]{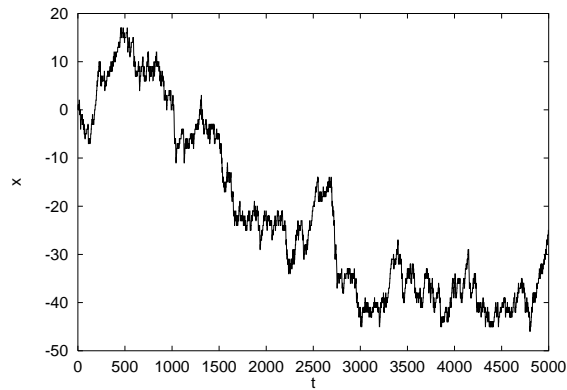}}
\caption{\label{fig:trace34}\textit{Random walker evolving according to Rule
6769347793221 (no 34).}}
\end{figure}

\begin{remark}As mentioned above, conjugation exchanges the roles of
particles and holes. Therefore, if a rule $f$ is self-conjugate, we can
always choose the rule radii $(r_\ell,r_r)$ such that the
point $(0.5,0)$ is a center of symmetry of its flow diagram. Rules
whose reference numbers are 26 and 34 are self-conjugate and their
flow diagrams have this property. In~\cite{BF} we studied a few 2-state
self-conjugate number-conserving rules. Some of these rules, which
allow motion in both directions, govern the dynamics of ensembles
of one-dimensional pseudo-random walkers. Starting from a random
initial configuration, whose density $\rho$ is exactly equal to
0.5, we followed, in the limit set, which is reached after a maximum
of $L/2$ iterations, the position of a specific particle as a function of
time on a ring of length $L=5000$ for 5000 time steps. The walks are
represented in Figures~\ref{fig:trace26} and \ref{fig:trace34}.
Since $L$ is finite, the random walks are periodic
in time. The first walk has a period equal to $L$, while the
period of the second one is $2L$. Note that these walks are deterministic,
their random character comes from the randomness of the initial configuration.
\end{remark}


\section{Conclusion}

We have established a necessary and sufficient condition for a
$q$-state $n$-input CA rule to be number-conserving. For given values of
$q$ and $n$, this result allows to find all the rules possessing
this property. As an example, we have determined all the three-state
three-input number-conserving CA rules. We have listed their motion
representation and studied their flow diagrams. All the rules
for which the flow is non-negative for all values of the average
density, can be considered as deterministic one-way road car traffic
rules in which the cells represent road sections whose car capacity
is equal to $q-1$. Among the 148 three-state three-input rules
there exist two nontrivial self-conjugate rules which mimic the
dynamics of an ensemble of random walkers.

\end{document}